**Abstract :** We used high order harmonics of a femtosecond titanium-doped sapphire system (pulse duration 25 fs) to realise Ultraviolet Photoelectron Spectroscopy (UPS) measurements on diamond. The UPS spectra were measured for harmonics in the range 13 to 27. We also made *ab initio* calculations of the electronic lifetime of conduction electrons in the energy range produced in the UPS experiment. Such calculations show that the lifetime suddenly diminishes when the conduction electron energy reaches the plasmon energy, whereas the UPS spectra show evidence in this range of a strong relaxation mechanism with an increased production of low energy secondary electrons. We propose that in this case the electronic relaxation proceeds in two steps : excitation of a plasmon by the high energy electron, the latter decaying into individual electron-hole pairs, as in the case of metals. This process is observed for the first time in an insulator and, on account of its high efficiency, should be introduced in the models of laser breakdown under high intensity.

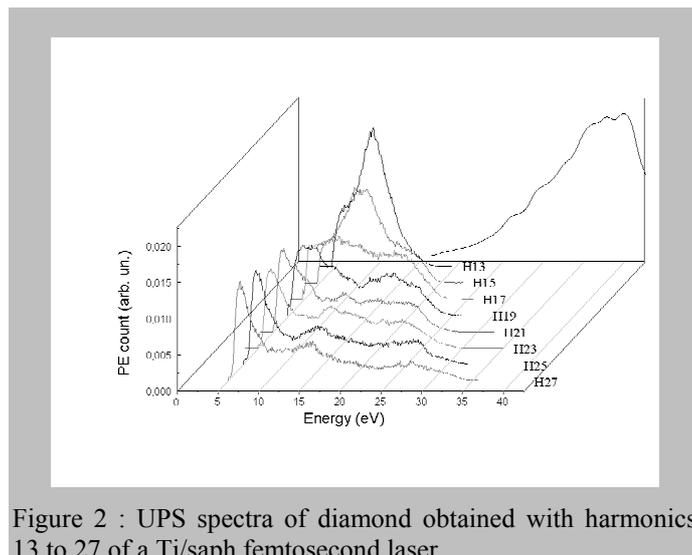

Figure 2 : UPS spectra of diamond obtained with harmonics 13 to 27 of a Ti/saph femtosecond laser.

# Plasmon channels in the electronic relaxation of diamond under high-order harmonics femtosecond irradiation.

*J. Gaudin* [(1)], *G. Geoffroy* [(1)], *S. Guizard* [(1)], *V. Olevano* [(1)], *S. Esnouf, S.M. Klimentov* [(2)], *P.A. Pivovarov* [(2)], *S.V. Garnov* [(2)], *P. Martin* [(3)], *A. Belsky* [(3)] *and G. Petite* [(1)*]

(1) Laboratoire des Solides Irradies, UMR 7642, CEA/DSM, CNRS/SPM et Ecole polytechnique, F-91128 Palaiseau Cedex, France
(2) General Physics Institute of the RAS , 38 Vavilov St., 119191, Moscow, Russia
(3) Laboratoire CELIA, UMR 5107, Universite Bordeaux I, 351 Cours de la Liberation, F-33405 Talence

**Key words** : electronic relaxation, laser breakdown, photoelectron spectroscopy, diamond

## 1. Introduction

The nature of the mechanisms of electronic relaxation is central to the problem of the breakdown of optical materials under high laser intensity. If in the nanosecond regime, there is a general agreement about a model following which, because the energy is slowly injected in the electronic system, it can be transmitted to the phonon system, the electrons basically staying in thermal equilibrium with the lattice. The situation is more intricate in the case of femtosecond irradiation. It is known that in this case transient out-of-equilibrium situations are common, with electrons reaching energies of tens of eV, the lattice still being at room temperature [1,2]. Such « hot » electrons can redistribute their energy within the electronic system, and there is still a living debate concerning the contribution of the avalanche mechanism to the free-electron density [3,4] induced in the material by the laser excitation, the answer depending on the choice of the threshold energy for « impact ionisation » and the type of model used (rate equations or full treatment of the collision integrals in a formalism based on the Boltzmann equation).

This issue of electronic relaxation channels is not specific to the problem of laser excited materials. It is at the base of a spectroscopy known as « Electron Energy Loss Spectroscopy » (EELS), and is for instance also central to electron microscopy of insulators. Thus a large amount of information is available there, that has been barely considered by the laser community, with the exception of the pioneering work of the IBM group [5 and references therein] on electron transport in SiO₂ layers. The question we would like to investigate in this paper is that of the plasmon excitation as a collective mechanism of the electron energy loss. Indeed, it is known to be a significant (if not dominating) energy loss channel for high energy electrons, and surprisingly enough, it is never considered in laser-solid interaction models, the interaction models taking into account only the individual excitation of secondary e-h pairs.

Quite obviously, one should be cautious not to directly apply information derived from EELS, which uses high energy electrons (tens of keV, if not hundreds), for which a 10 eV (individual e-h excitations) to 50 eV (plasmons) energy loss is negligible, to electrons with a few tens of eV of energy (the case of a femtosecond laser excitation, close to the Optical breakdown Threshold – OBT) which may be able to excite plasmons, but will practically loose all their energy doing so. Therefore the control of the excitation spectrum is essential, and direct use of a femtosecond laser is not the best tool for this purpose. In the following, starting from the fact that titanium doped sapphire excitation of insulating samples at intensities in the range of TW.cm$^{-2}$ produces electrons with typically 30 eV of maximum kinetic energy [6], we used as an excitation source the high order harmonics of such a laser, which can produce electrons with similar kinetic energies, but whose spectrum can be controlled by changing the order of the harmonics. The energy of free electrons resulting from such an excitation is monitored using photoelectron sepctroscopy. We present here first results obtained on diamond, a material that was selected both because of its important applications to detection of ionising radiation [7] – which of course involves electron transport – but also because the first *ab initio* calculations of the electronic lifetimes – which is determined, for its electronic part, by the energy loss mechanisms we discussed above – have been realised, and will be presented below.

## 2. Lifetime calculations

Theory of excited electronic states recently made significant progresses [8], so that it is now possible to obtain through *ab initio* calculations theoretical values of the lifetime of electronic excitations, as determined by the different types of electron-electron interactions that we mentioned above. In practice, one starts from the determination of the quasi-particle (the correct designation of the "conduction electrons" that we mentioned above) self-energy operator expressed in the "GW" approximation [9]

$$\Sigma^{GW}(r_1, r_2, \omega) = \frac{i}{2\pi} \int d\omega' G^{DFT-LDA}(r_1, r_2, \omega-\omega') W^{RPA}(r_1, r_2, \omega') \quad (1)$$

where $G$ is the quasi-particle Green's function and $W$ a screened Coulomb interaction (in eq. (1) and hereafter, the upper indices express the approximation in which the quantities are evaluated - 19 special k-points in the irreducible Brillouin zone are used for the evaluation of these quantities, which must include high symmetry points). One calculates the integral (1) using a Gauss-Legendre quadrature, with 30 Gauss knots. The self-energy is then evaluated in a number of discrete points on the imaginary frequency axis (12 points on a linear mesh between 0 and 200 eV). An analytical continuation of $\Sigma^{GW}$ into the whole complex plane is then obtained using a Pade approximant $P(\omega)$, since one is interested in quasi-energies that lie close to the real axis. The interest of starting from the imaginary axis is that the function $\Sigma^{GW}(\omega)$ is more regular there. The quasi-particle energy is then obtained by solving the equation

$$E_{QP} = E_{KS} - E_{xc}^{LDA} + P(\omega = E_{QP}) \quad (2)$$

where $E_{KS}$ and $E_{xc}$ are respectively the Kohn-Sham and exchange-correlation energies. To do so, one uses the Newton-Raphson method (starting from $\omega=E_{KS}$), an iterative method, which is stopped when the precision reached is better than 0.001 eV. Finally, the quasi-particle lifetime is calculated as

$$\tau^{-1} = 2\Im(E_{QP}) \quad (3)$$

Since we are neglecting the ionic degrees of freedom in the self-energy (1), the resulting lifetime accounts only for electronic decay channels (e.g. decay by emission of plasmons, e-h pairs, Auger processes). Phonon emission decay and scattering with defects and impurities are neglected. Therefore the lifetime is meaningful only where electronic processes dominate, i;e; far away from the Fermi level. Figure 1 shows the results of such a calculation in the case of diamond, where we concentrated on the 10-50 eV region, which is of interest here (the origin of energies is taken at the top of the valence band), because electronic scattering dominates.

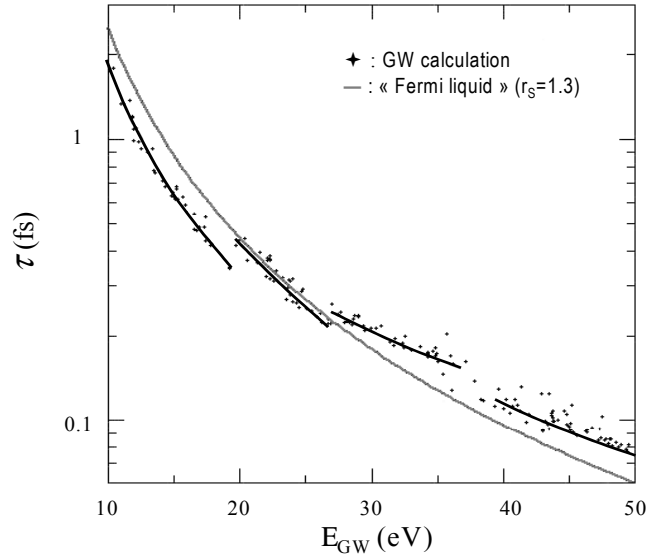

Figure 1 : GW electronic lifetime of excited electrons in diamond. The continuous curve shows for comparison the prediction of a "Fermi liquid" type calculation, using the electronic density of diamond. The black lines are added by hand as a guide to the eye.

The results are compared with the predictions of a Fermi liquid model using the parameters of diamond, with which they show significant differences, but staying within the same order of magnitude. The purpose of this figure is to make two remarks :

- the electronic lifetime is very short, typically 0.5 fs at a 10 eV of energy in the conduction band (bandgap subtracted from GW energies of fig. 1). This allows a very rough estimation of the electron mean free path (using free-electron parameters) which is of the order of the nm. Using the more correct parameters (which requires to enter into the details of the diamond band structure) one can assume that this quantity will somewhat increase (the electron effective mass is smaller than unity) but by less than one order of magnitude. In the following experiment, we will use harmonics whose penetration depth ranges from typically 5 to 15 nm. We can then safely assume that many of the electrons composing the photoemission spectrum (but clearly not all of them) will have been subject to at least one interaction. One thus expect the electron spectra to be a combination of the direct excitation spectrum and of relaxed electrons.
- the general behaviour is not very different from the Fermi liquid one, but some differences are clearly visible. We can identify four regions with a rather regular behaviour, with three transition zones : at 20 eV, a sudden increase (typically 20%) of the lifetime, and a somewhat smaller at 26 eV, which we attribute to band structure effects. More important for us here is the rather rapid decrease (by about 30%) observed around 35 eV, which, once the bandgap energy retrieved, corresponds almost exactly to the diamond Drude plasmon energy (31.5 eV). In fact this transition region exactly coincides with the plasmon peak found in an RPA calculation of the EELS spectrum, but one should remember that such a calculation is made for a high electron initial energy, so that it does not directly apply to the situation we investigate. But the lifetime calculation does, and it also shows an quite significant effect in this region.

Let us finally note that these features only appear if, among the 19 k-points employed for the calculation, we include high-symmetry points, a sign that symmetry of diamond is of importance here, which deserves some further investigations.

## 3. Experimental setup and results.

The photoelectron energy spectra (UPS) of diamond were measured using as an excitation source harmonics 13 to 27 of the CELIA laser (in University Bordeaux-I). The sample was a type IIa single-crystal of high purity, with dimensions 5x5x0.5 mm$^3$. Prior to the experiment, its defect content was checked using EPR spectroscopy, which revealed very few dangling bonds and practically no P1 centres, meaning a very low content of nitrogen.

The harmonics were selected using a low resolution grating, which allowed to minimise the unavoidable pulse lengthening. The pulse duration, starting from the 25 fs corresponding to the laser, is estimated to about 1 ps. The repetition rate was 1 kHz

The sample was placed in a UHV vessel, and irradiated by the harmonics using a p-polarisation. It was heated to a temperature of about 300°C, in order to provide enough conductivity to allow compensation of the charge left by the emitted electrons between two laser shots, which permits to cancel the energy shifts otherwise observed during the acquisition of one spectrum. The spectrometer used was a CLAM-IV hemispherical electrostatic spectrometer (mean radius : 150 mm), with a wide aperture (12°) collection optics. A 25 eV pass energy was employed, which corresponds to an energy resolution of less than 0.1 eV. Figure 2 shows the photoelectron spectra measured in the experiment. They have been normalised to a constant number of electron in the different spectra. Because of the usual contact potential problems (uncertainty in the position of the Fermi level of the sample compared to that of the spectrometer), the spectra shown on fig. 2 were reset in order to position their low energy edge at the theoretical energy of the bottom of the conduction band (5.5 eV). In this procedure, we also compensated a small (close to 1 eV) general energy shift which was observed between the first spectra acquired and the last ones, probably corresponding to a global residual charging of the sample (but was always negligible over the acquisition time of one single spectrum).

## 4. Discussion and conclusions

The photoelectron spectra of figure 2 show a quite noticeable evolution in their shape when the harmonics index increases. There is also some amplitude evolution, but it mainly reflects the evolution of the ratio between penetration depth of the radiation and escape depth of the electrons, and would require for its discussion a realistic calculation of both the excitation spectrum and the transport to the surface, both possible but beyond the scope of this paper. We thus concentrate hereafter on the main qualitative features.

The spectrum obtained with harmonic 13$^{th}$ essentially shows a main peak around 10 eV and a low number of band edge electrons. Let us point out here that there is no simple hand-waving argument allowing to decide of the structure of a spectrum like this one when one deals with complicated band structures (which is the case here). The calculation is feasible but not yet available. We just note that 10 eV corresponds typically to the centre of the first conduction band of diamond, the only one which can be excited in this case [10]. We also note that the photon energy is here of 20.5 eV, yielding maximum energies in the CB for the excited electrons of 15eV, in principle large enough to allow secondary e-h pairs excitation. Obviously this does not affect significantly the spectrum. One could understand this as an effect of the low penetration depth (so that the probability for an electron to escape without interacting with the others is still high) and also on account of the fact that only a small number of the excited electrons (those excited above the threshold energy, that is here typically between 1.5 and 2 times the bandgap above the minimum of the CB) are able to excite secondary e-h pairs. The spectrum obtained with H15 roughly presents the same structure, except for a high energy tail which naturally results from the increase of the photon energy (and corresponding to excitation of states in a second conduction band of diamond, separated from the first one by a small bandgap close to 1 eV [11]). The peak at 10 eV is still clearly visible. On the contrary, the spectrum obtained with H17 appears almost flat, the feature at 10 eV being still visible, but attenuated. The low energy electrons, whose contribution appears stronger can originate from relaxation processes, but in the case of H17, can still be obtained by direct excitation of states in the bottom of the valence band.

Some other experimental evidence is available concerning the generation of secondary e-h pairs in this photon energy range, from results recently obtained of the pulsed photoconductivity induced in diamond by high order harmonics [12]. This method allows to measure the efficiency of a femtosecond harmonic pulse in terms of total e-h pair excitation, and thus to detect an eventual multiplication mechanism. It was shown that this efficiency increases in diamond from H9 to H13 (which would be in accordance with the individual secondary pair excitation mechanism) but then decreases from H13 to H17. This is a proof that the mechanism of secondary pair creation is more complicated that the simple model usually used in the different laser-breakdown models, and it was suggested that band structure effects could play a significant role [13], but some other possible explanations are also considered. Though the spectra of fig 2 cannot be directly compared to such photoconductivity measurements, they certainly bring no contradictory evidence (such as the appearance in this harmonic range of a significant secondary peak).

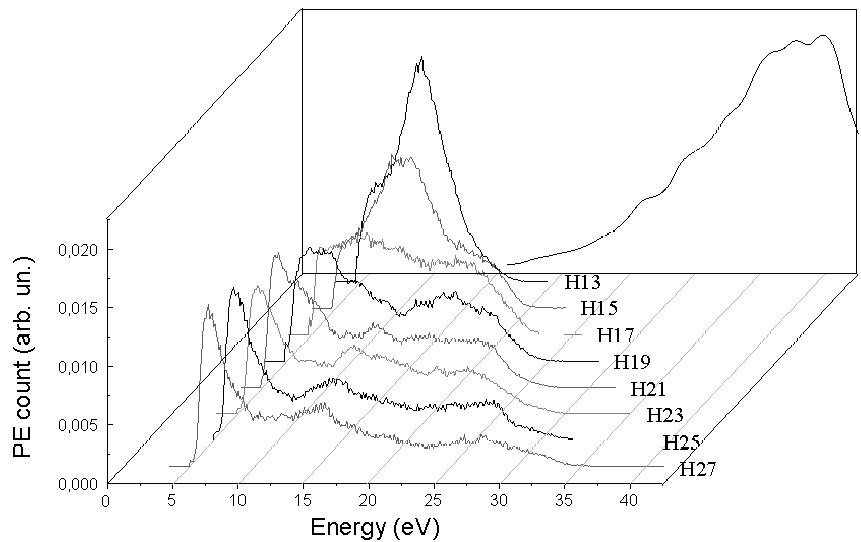

Figure 2 : UPS spectra of diamond obtained with harmonics 13 to 27 of a Ti/saph femtosecond laser. The spectrum on the back panel is the result of the EELS calculation mentioned in section 2.

Starting from H19 and up to H27, one observes a general behaviour which can be described as follows : appearance of a strong low–energy (secondary) peak, which is accompanied by an obvious deficit of high energy electrons, when the evolution of the spectrum extension is compared to that of the photon energy (the maximum possible energy is indicated by the end of the solid line). Let us note here that the simple fact that electrons are measured near the bottom of the CB is a clear evidence of some relaxation process. If one takes, e.g., H19 with a photon energy of 30.5 eV, it cannot excite electrons with an energy of less than 10 eV (on account of the 20 eV total width of the valence band). Let us point out that in diamond, the electron-phonon coupling is quite low (hence its excellent carrier mobility) so that the relaxation mechanism is most likely essentially of electronic origin.

From fig. 2, it follows that a quite efficient relaxation mechanism sets in in this energy range. It is remarkable that this occurs for a range of harmonics in which the maximum energy of the excited electrons is located in the region of the plasmon energy loss. We therefore consider Figure 2 as an evidence of the existence of an electronic relaxation mechanism which was so far disregarded in the framework of laser-solid interaction, by which secondary e-h pairs are created through an indirect channel : first, excitation of a plasmon (at the origin of the consumption of high energy electrons), which further decays into individual pairs (which form the low energy peak clearly visible on the corresponding spectra of fig. 2). Decomposition of plasmons in a sum of individual excitations is a known phenomenon in metals' physics [14] and it has no reason not to apply to the case of insulators.

Let us consider some arguments that could be raised against such an interpretation. A first one is that the above observations result from the rapid decrease of the electron mean free path in this energy range. This is in fact true, but it is just consequence of the electron scattering processes (whatever they are), so that this is not another interpretation, but only another way of depicting the physical processes we have been discussing. Another one is that the relaxation mechanism does not need to be collective, but could simply be a sequence of individual e-h pair excitations. This possibility should certainly be seriously considered, but there are a number of arguments against it : first this should occur for lower harmonics (certainly in the case of H17) and both the moderate evolution of the spectrum in this, and the results of ref [12] indicate that this process is not so efficient, and one does not see why the efficiency should increase while the number of elementary events increases. Second, lifetime calculation (see fig. 1), which only involve single events (plasmon or individual pair excitations) show that something is happening in the plasmon energy loss region. Finally the shape of the EEL function shows that plasmon excitation is much more efficient than single pair excitation (whose probability is typically represented by the curve amplitude in the 10 eV region), but this will be a definite proof only when such a calculation will be available for low energy electrons.

As a conclusion, we have brought a number of indications, both experimental and theoretical, that an electronic relaxation channel that was so far disregarded in the framework of laser-solid interactions (and particularly in that of laser breakdown of optical material under irradiation by high intensity femtosecond pulses) may be of importance, and even dominating. This relaxation channel consists in plasmon excitation by the high energy electrons, followed by the decay of this plasmon into low energy individual e-h pairs.

We have shown that photoelectron spectroscopy using high order harmonics, which allow to control the excitation spectrum, is a useful tool for investigating such questions, but also that the interpretation of the spectra is a complicated matter. Comparison of the experimental spectra with the results of *ab-initio* calculations of the excitation spectrum which are now possible would certainly greatly improve our understanding of the problem. Experimentally, measuring the photoconductivity induced by high order harmonics in the same range, which allow a measurement of the total bulk concentration of carriers induced by the excitation, would also be of great interest.

## Acknowledgements

The authors acknowledge the support of INTAS program n° 01-458 and of a Franco-Russian PAI (n° 04527 ZM). J.G. acknowledges the support or Region Aquitaine.